\documentclass[runningheads]{llncs}
\usepackage{graphicx}

\usepackage{color}

\usepackage{cite}
\usepackage{multirow}
\usepackage{algorithm}
\usepackage{algpseudocode}
\usepackage{amssymb}
\usepackage{gensymb}
\usepackage{wrapfig}

\usepackage{cleveref}

\begin{document}
\title{Are Sex-based Physiological Differences the Cause of Gender Bias for Chest X-ray Diagnosis?}
%
%
\author{Nina Weng\inst{1}
\and
Siavash Bigdeli\inst{1}
\and
Eike Petersen\inst{1}
\and
Aasa Feragen\inst{1}
}
\authorrunning{Weng et al.}
%
\institute{Technical University of Denmark, Denmark\\
\email{\{ninwe,sarbi,ewipe,afhar\}@dtu.dk}}
\maketitle              

\vspace{0.5cm}

\begin{abstract}
While many studies have assessed the fairness of AI algorithms in the medical field, the causes of differences in prediction performance are often unknown. This lack of knowledge about the causes of bias hampers the efficacy of bias mitigation, as evidenced by the fact that simple dataset balancing still often performs best in reducing performance gaps but is unable to resolve all performance differences. In this work, we investigate the causes of gender bias in machine learning-based chest X-ray diagnosis. In particular, we explore the hypothesis that breast tissue leads to underexposure of the lungs and causes lower model performance. Methodologically, we propose a new sampling method which addresses the highly skewed distribution of recordings per patient in two widely used public datasets, while at the same time reducing the impact of label errors. Our comprehensive analysis of gender differences across diseases, datasets, and gender representations in the training set shows that dataset imbalance is not the sole cause of performance differences. Moreover, relative group performance differs strongly between datasets, indicating important dataset-specific factors influencing male/female group performance. Finally, we investigate the effect of breast tissue more specifically, by cropping out the breasts from recordings, finding that this does not resolve the observed performance gaps. In conclusion, our results indicate that dataset-specific factors, not fundamental physiological differences, are the main drivers of male--female performance gaps in chest X-ray analyses on widely used NIH and CheXpert Dataset.

\end{abstract}
\section{Introduction}
AI fairness receives increased attention with the escalating demand for examining the validity and responsibility of AI methods. This is particularly crucial in the medical field, where automatic and intelligent decision-making algorithms could easily lead to unfair treatment without the awareness of fairness.

\begin{wrapfigure}{r}{0.4\linewidth}
\vspace{-0.7cm}
\centering
\includegraphics[width=1\linewidth]{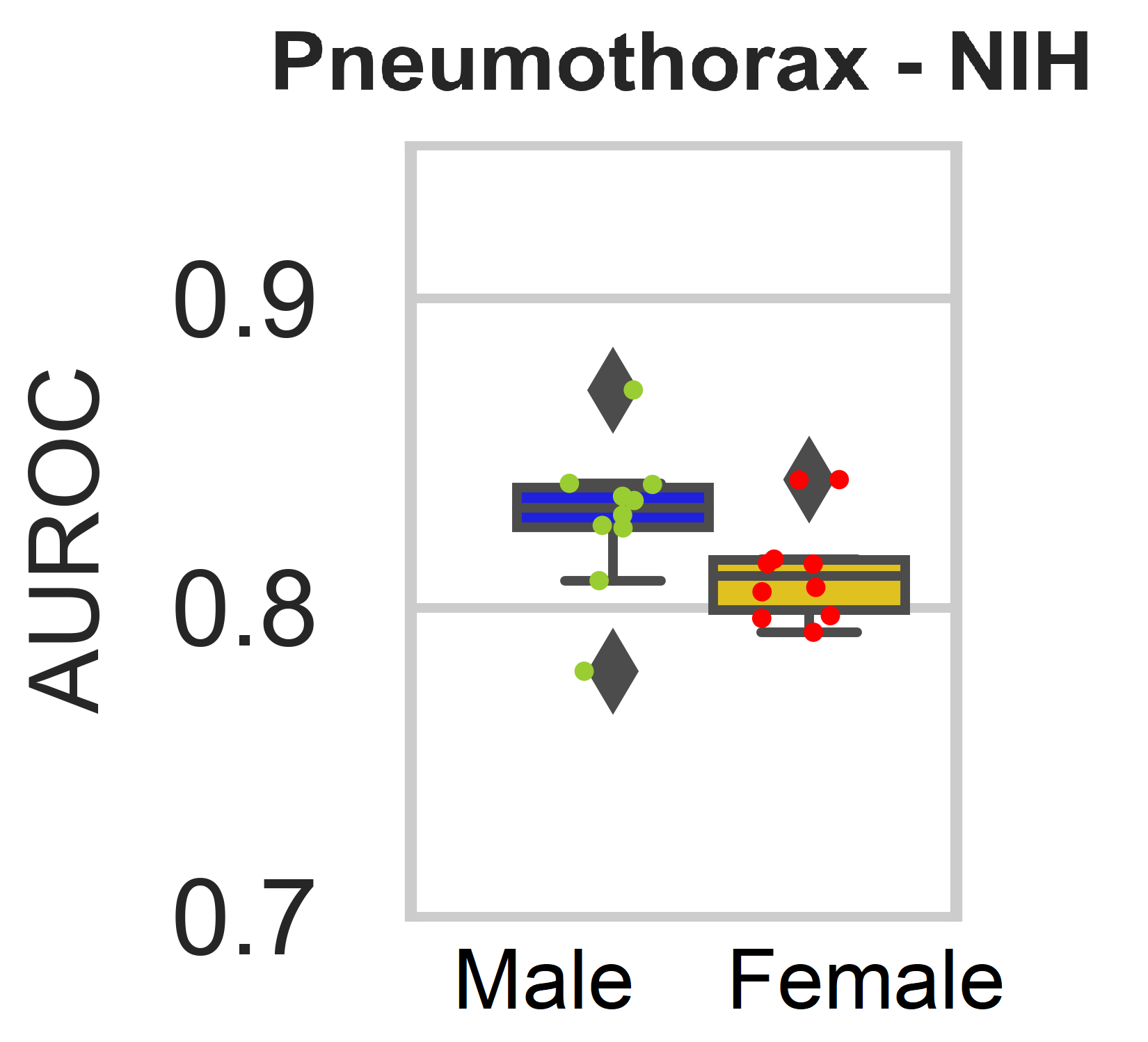}
\caption{An excerpt of our results, inspired by~\cite{larrazabal2020gender}, showing Pneumothorax diagnosis performance evaluated on men and women for an algorithm trained \emph{solely} on women.
In this example, even female over-representation does not yield equal performance. This example led to a hypothesis that female breasts might lead to degraded image quality~\cite{ganz2021assessing}.
}
\label{fig:motivation}
\vspace{-0.6cm}
\end{wrapfigure}

A series of studies have assessed fairness in various medical imaging settings, including chest x-rays~\cite{seyyed2020chexclusion, larrazabal2020gender}, retinal imaging~\cite{burlina2021addressing}, brain MRI~\cite{stanley2022fairness, petersen2022feature}, and cardiac MRI~\cite{puyol2022fairness}. While different types of bias mitigation techniques have been applied~\cite{Daneshjou2022, Wu2022, zhang2022improving, Pakzad2023}, there is currently very limited work that seeks to diagnose the \emph{causes} of bias~\cite{Petersen2023a}, enabling the targeted selection of bias mitigation methods. Supporting the urgency of this type of \emph{bias reasoning}, ~\cite{zhang2022improving} shows, by comparing the performance of different bias mitigation strategies, that simply balancing datasets is still one of the best strategies for mitigating gender bias for chest X-ray diagnosis. 
At the same time, studies have shown that while the level of group representation affects group-wise performance, balancing datasets alone does not guarantee equal performance across groups~\cite{larrazabal2020gender}, see \cref{fig:motivation} for an example. This implies that without further investigation of the causes of bias, our attempts to mitigate bias might be very limited in success, as also evidenced by the well-known `leveling down' phenomenon~\cite{Zietlow2022}.

In this paper, we investigate causes of gender bias in machine learning-based chest X-ray diagnosis, where significant performance disparities between genders are observed.
Previous works~\cite{ganz2021assessing,chestradiologykey, jenkins2013making, alexander1958elimination} have suggested that breast tissue might lead to impaired image quality, and hence lower performance, of chest X-ray-based diagnostic classifiers. 
Here, we perform a series of experiments to analyze whether the observed gender bias is indeed a result of physiological sex differences. 
In short, our contributions include:
\begin{enumerate}
\item \textbf{A new way of sampling} training and test sets from publicly available chest X-ray datasets 
that reduces the influence of potential confounders, such as a highly skewed distribution of the number of recordings per patient and missing disease labels, on training and analysis. In particular, we propose to sample just a single recording per patient, preferring samples with a disease-positive label.
A comparison of different sampling strategies provides further strong evidence of label errors in publicly available datasets.
\item A combination of our proposed sampling method with training sets of varying gender ratios to provide \textbf{a comprehensive re-analysis of gender differences} in model performance across multiple diseases in two well-known datasets (CheXpert~\cite{irvin2019chexpert} and NIH ChestX-ray8~\cite{wang2017chestx}).
Our results indicate that imbalanced datasets are not the only cause of performance differences, and, crucially, that gender-based performance differences differ between datasets even for the same disease.
\item Further \textbf{experiments designed to study whether female breasts cause diagnostic classifier bias}. We assess how cropping out the breasts from recordings affects model performance, and we consider inter-dataset transfer of models for more evidence of inherent gender bias from the dataset.
We find that female breasts do not appear to be a strong cause of model performance disparities and that there appear to be other, presently unknown dataset-dependent factors influencing model performance.
\end{enumerate}

\section{Related Work}
Following claims of radiologist-level performance in machine learning-based chest X-ray disease classification~\cite{Rajpurkar2017}, performance disparities of such disease classifiers between patient groups have come under increased scrutiny~\cite{larrazabal2020gender,seyyed2020chexclusion,SeyyedKalantari2021}.
Larrazabal et al.~\cite{larrazabal2020gender} analyzed the effect of gender imbalance in training sets on performance disparities in the trained classifier, finding a strong link between the two.
Moreover, their results show how performance in some groups remains poor even if models are trained \emph{only} on subjects from that group; refer to \cref{fig:motivation} for an example.
Seyyed-Kalantari et al.~\cite{seyyed2020chexclusion,seyyed2021underdiagnosis} showed that state-of-the-art classifiers consistently and selective underdiagnosed historically underserved patient populations, such as non-white and female patients.
Later, Zhang et al.~\cite{zhang2022improving} investigated a range of possible bias mitigation techniques, finding that simple group balancing still appears to be the most successful mitigation technique; this result has also been confirmed in other contexts~\cite{Idrissi2022}.
Taken together, these results emphasize the importance of well-representative datasets.
However, as Larrazabal et al.~\cite{larrazabal2020gender} had shown, group balancing alone cannot alleviate all performance differences, thus emphasizing the urgent need for more nuanced investigations into the \emph{sources} of bias to enable successful bias mitigation.

In the wake of these important studies, there have been several investigations into potential sources of bias in chest X-ray datasets.
In response to the study of Seyyed-Kalantari et al.~\cite{seyyed2021underdiagnosis}, Bernhardt et al.~\cite{bernhardt2022} and Glocker et al.~\cite{Glocker2023} pointed out the importance of properly accounting for confounding factors in bias analyses, 
such as age and disease distributions between patient groups.
Bernhardt et al.~\cite{bernhardt2022} moreover underlined the challenge of properly evaluating performance differences if \emph{label biases} affect both the training and test sets.
There is reason for concern in this regard, since multiple studies have reported high error rates in (NLP-derived~\cite{irvin2019chexpert}) chest X-ray disease labels~\cite{zhang2022improving,smit2020}.
Our results in this study provide further independent evidence of widespread label errors in these databases.

Separately from these methodological issues, the performance differences between male and female patients led Ganz et al.~\cite{ganz2021assessing} to speculate that an important cause might lie in female breasts occluding the recordings of important lung regions.
Indeed, the confounding effect of female breasts on clinical chest x-ray interpretation is well-known~\cite{chestradiologykey, jenkins2013making, alexander1958elimination} and physiologically plausible: additional breast tissue results in the underexposure of lung tissue.
In this work, we take a step to systematically assess the effect of female breast tissue on machine learning-based chest X-ray diagnosis.

\section{Methods}

\subsection{Datasets}

We consider two datasets: ChestX-ray8 (NIH)~\cite{wang2017chestx} and CheXpert~\cite{irvin2019chexpert}. As the NIH dataset only contains frontal images, we also only use those views from the CheXpert to enable a fair comparison, resulting in 112,120 recordings from 30,850 patients in the NIH dataset and 190,299 recordings from 64,224 patients in the CheXpert dataset.
Both datasets slightly over-represent male subjects (54\% males vs. 46\% females in NIH, 56\% males  vs. 45\% females in CheXpert); refer to 
table~1 in the supplementary material for further details.

\subsection{Sampling Strategy}
\label{ssec:sampling}
\subsubsection{Motivation.}
We observe that in both datasets, \emph{the number of recordings per patient is very uneven}, ranging from 1 to 89 (CheXpert) and 1 to 184 (NIH). In particular, 
less than 25\% of controls and over 50\% of patients have more than 5 scans. 
This results in few patients with many recordings strongly influencing the training process and the final model, as well as strong distribution shifts between different data splits.

In addition, like outlined above, it has been shown that \emph{disease labels automatically derived from patient records are unreliable}~\cite{zhang2022improving}.
Particularly, it has been observed~\cite{lauranchestxray} that the commonly used text mining method worked poorly with the hospital record of ``no change from previous'', which would be wrongly marked as “no finding”. 
Labels might be especially unreliable in patients with many recordings.

For these reasons outlined above, we select only one sample from each patient to avoid over-representation with a preference for positive labelled samples.

\subsubsection{Principles underlying the proposed sampling strategy.} \label{sec:principle}
We designed our sampling strategy to conform with the following principles:
\begin{itemize}
    \item Utilize one sample per patient: reducing risk of distribution shift between splits and over-reliance on individual subjects. 
    \item Prioritize diseased samples when selecting the single sample per patient: reducing risk of label bias towards ``no finding'' like outlined above. 
    \item Keep disease prevalence constant across splits: 
    reducing risk of distribution shift between splits.
    \item Allow disease prevalence to vary between protected groups: ensuring that our assessment is realistic, by utilizing realistic prevalences across groups.
    \item For training and validation, draw a fixed-size sample with a predefined percentage of female subjects (0\%, 50\% or 100\%): enabling an assessment of the influence of training set composition on model performance. (For testing, we always draw a fixed-size sample with an equal number of males/females.)
    \item Use an identical test set when evaluating the same training split at different gender ratios: enabling more reliable performance comparisons.
\end{itemize}

\noindent For further details on our sampling scheme, refer to Algorithm~1 in the supplementary material.

\subsection{Experimental Settings}
All experiments are carried out using the PyTorch framework with a pretrained ResNet50~\cite{he2016deep} and the Adam optimizer~\cite{kingma2014adam} (learning rate~$10^{-6}$, 20 epochs, batch size~64).
We split the datasets into 60\% / 10\% / 30\% train, validation, and test sets per gender ten times
based on disease prevalence,
and then train separate single-label classifiers for all disease labels in both datasets.
During training, data augmentation is applied by random horizontal flipping, rotation (degree $\leq$ 15$\degree$) and scaling (from 0.9 to 1.1) with a probability of 0.5 each. 
Following~\cite{larrazabal2020gender}, experiments are run on three different gender ratios, i.e. 0\%, 50\% and 100\% females.
Performances are evaluated by the area under the receiver-operating characteristic curve (AUROC).

\section{Results}

\subsection{Model performance across diseases, gender ratios, and datasets}
\label{ssec:model-performance}
\begin{figure}[t]
\centering

\includegraphics[width=1.0\textwidth]{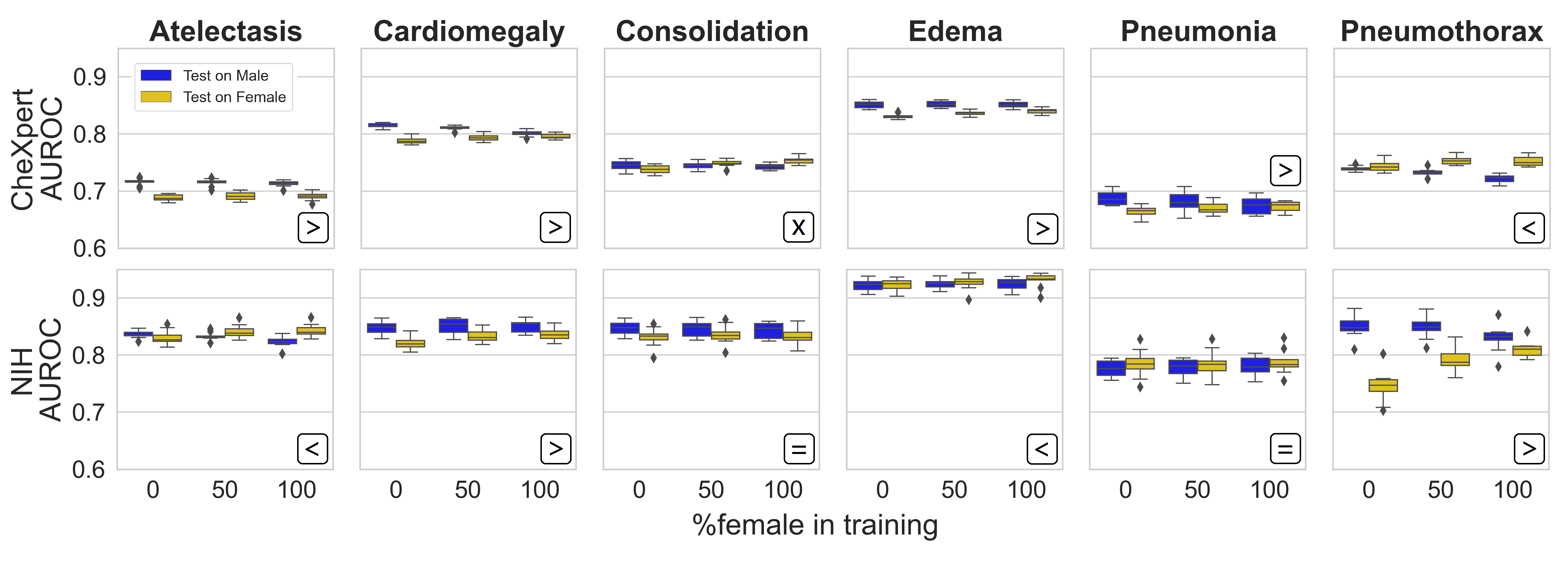}
\caption{Model performance (AUROC) in male (blue) and female (yellow) test subjects for training datasets with differing gender representation. 
Results are only shown for the six disease labels that are present in both datasets; for results on the remaining labels refer to the supplementary material.
Icons in the bottom right of each plot indicate the observed performance trends across gender ratios; refer to Section~\ref{ssec:model-performance}.
}
\label{fig:all_per}
\vspace{-0.5cm}
\end{figure}

\Cref{fig:all_per} displays male and female test subject performance under varied in-training gender ratios for the common 6 disease labels in both datasets.
Moreover, a stylized summary of the observed trends across gender ratios (`$\times$', `$>$', `$<$', or `$=$') is marked in the figure.
We highlight three key observations.

\paragraph{Dataset imbalance is not the sole cause of performance differences.} 
In some dataset--disease combinations, such as Pneumothorax--NIH, males keep outperforming females regardless of the proportion of women in the training set.
Similar and opposite trends (males having worse performance than females across gender ratios) could be observed in other disease labels, marked by `$>$' and `$<$' in \cref{fig:all_per}.
If performance differences were caused solely by training set imbalance, then the majority group should consistently outperform the minority group (resulting in an `$\times$' trend shape), which is not observed in most of the cases.

\paragraph{Performance trends sometimes differ strongly between datasets for the same diseases.} 
Consider again the case of Pneumothorax classification, and compare the results between the two datasets: 
On NIH, males demonstrate significantly better performance regardless of training gender distribution, while on CheXpert, the trend has reversed completely.
This suggests that there are dataset-specific factors unrelated to fundamental biological sex differences that strongly influence performance disparities.

\paragraph{Higher performance and higher variance on NIH compared to CheXpert.} 
Across all six shared disease labels, test-set prediction performance is considerably better on NIH compared to CheXpert, despite the smaller NIH training set size.
At the same time, the NIH performance also shows a larger variance across diseases, which appears likely to be related to the smaller test set size.

\subsection{Comparison of different sampling strategies}

\begin{figure}[t]
\centering
\includegraphics[width=0.95\textwidth]{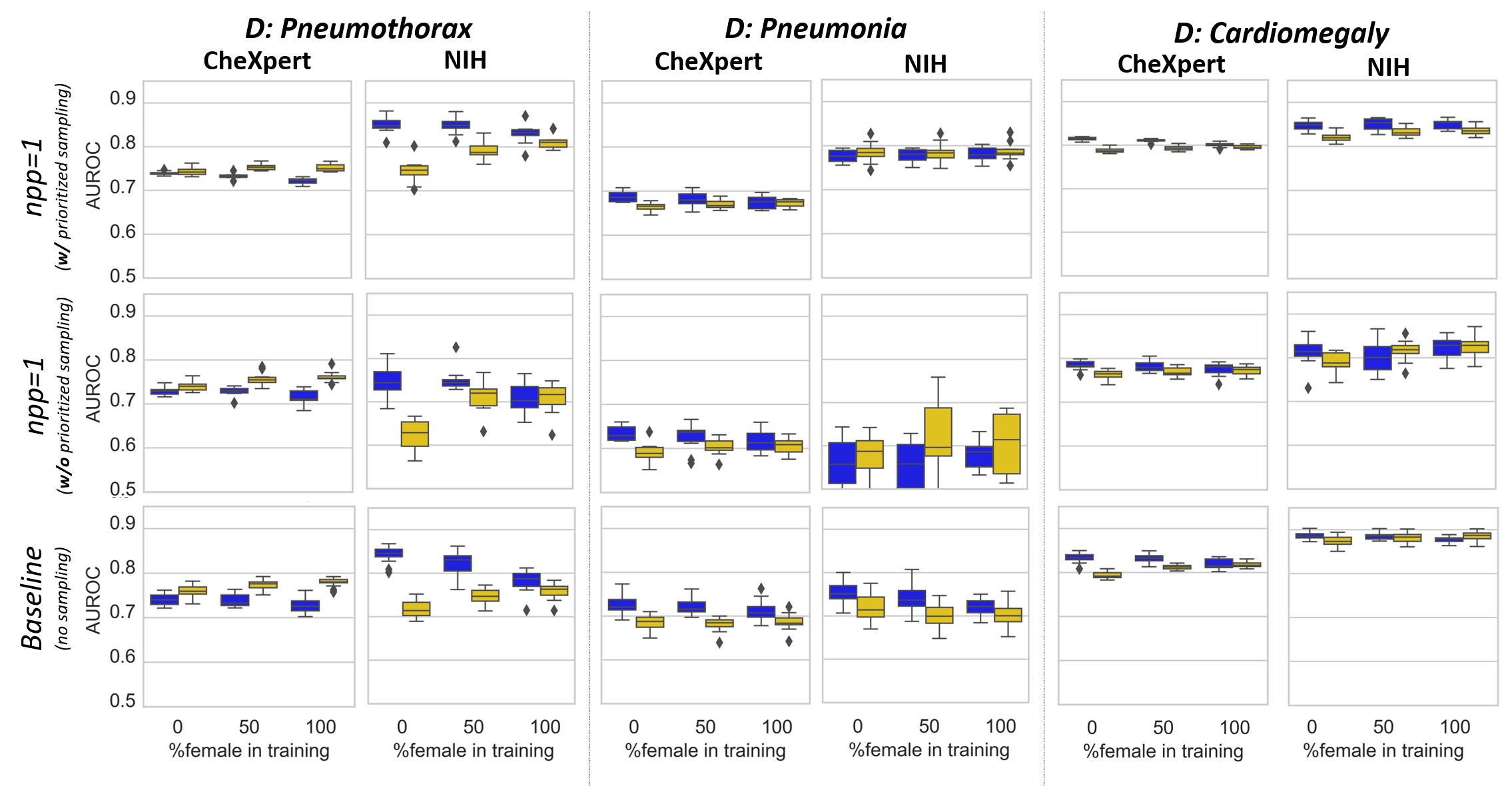}
\caption{Comparison of results from 10 varied splits when training and testing using three different sampling strategies for three disease labels. Three rows present one sample per patient (\texttt{npp=1}) with/without prioritizing diseased samples, and the original setting from \cite{larrazabal2020gender} which uses all the samples with multi-label learning. Blue box represents male performance and yellow refers female.
}
\label{fig:com_between_setups}
\end{figure}

\begin{figure}[ht]
\centering
\includegraphics[width=0.7\textwidth]{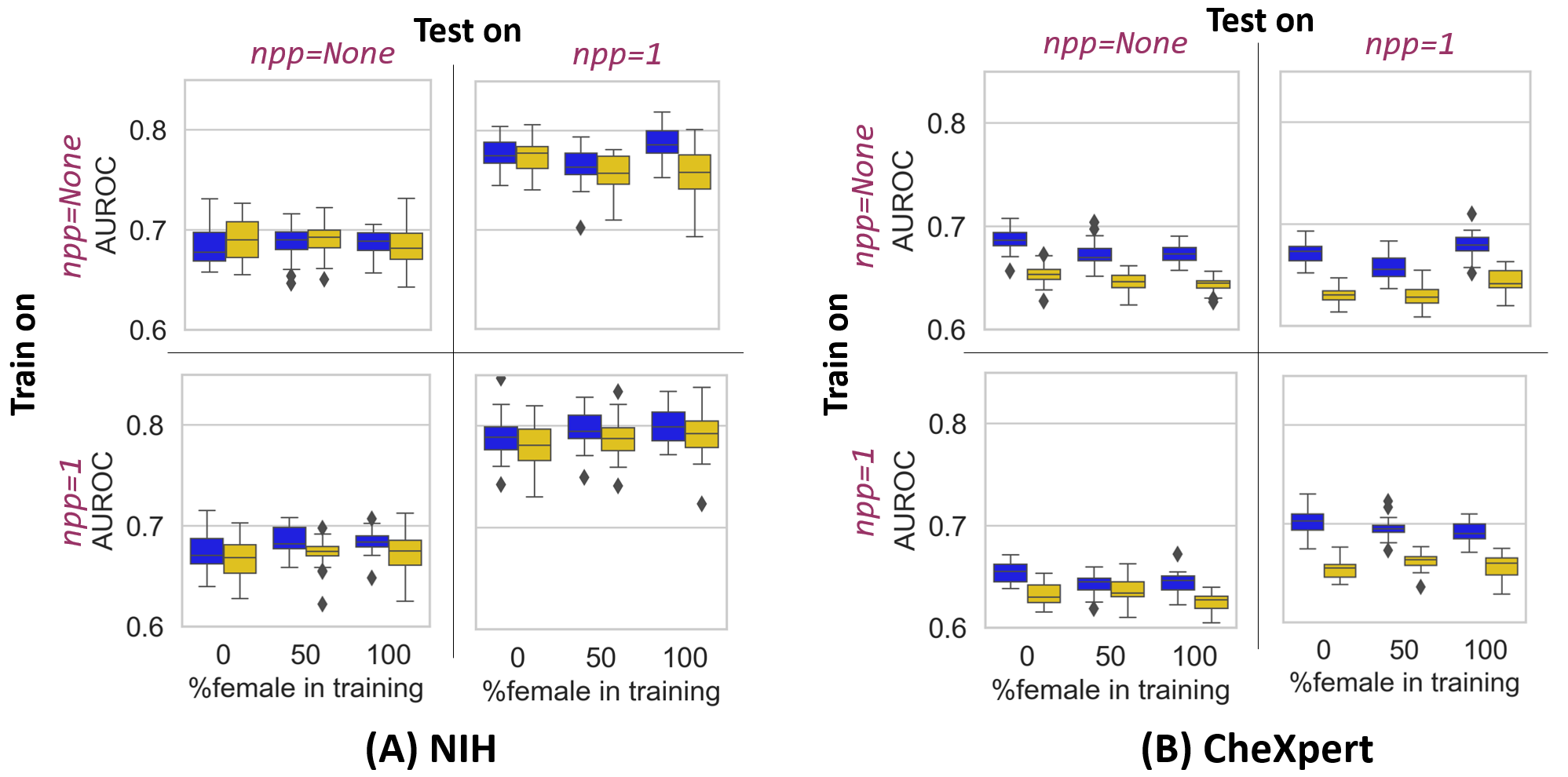}
\caption{Comparison of results when drawing the training set according to one sampling strategy and the test set according to another, for the disease label Pneumonia.
Error bars computed from a single training run using test set bootstrapping.
}
\label{fig:com_across_setups}
\end{figure}

To analyze the effect of our proposed sampling strategy, we first compare the results when training and evaluating using three different sampling strategies: 1) one sample per patient, preferring disease-positive samples (our proposed strategy), 2) one sample per patient, drawing the sample randomly and not preferring disease-positive samples, and 3) using all samples per patient with the same split and multi-label training as~\cite{larrazabal2020gender}. 
Note that in the latter case, we use larger training set sizes.
As shown in \cref{fig:com_between_setups},
in both datasets, test set performance is consistently and considerably improved when using the disease-preferring sampling strategy over a random sampling of one recording per patient;
we interpret this as evidence of strong label noise in the ``no finding'' label like discussed in \cref{ssec:sampling} and like reported previously in other datasets automatically labeled using the CheXpert labeller~\cite{zhang2022improving}.
The difference between the two setups is especially strong in the NIH dataset, leading to the hypothesis that NIH dataset might suffer from more widespread label errors compared to CheXpert.

Since label noise confounds not only training but also evaluation, we also present results for one disease when the training on one strategy and the test set according to another (\cref{fig:com_across_setups}). Due to the potential data leakage through train and test set, instead of using the splits from~\cite{larrazabal2020gender}, we implemented the experiments with all samples (\texttt{npp=None}) using modified proposed strategy without sampling.
The results indicate label errors are more likely occur in NIH than CheXpert, as the results in CheXpert are more stable across sampling strategies while the performance drops when testing on all samples in NIH.

\subsection{Breast cropping does not mitigate gender biases} \label{sec:cropped}

\begin{wrapfigure}{r}{0.4\linewidth}
\vspace{-1.0cm}
\centering
\includegraphics[width=\linewidth]{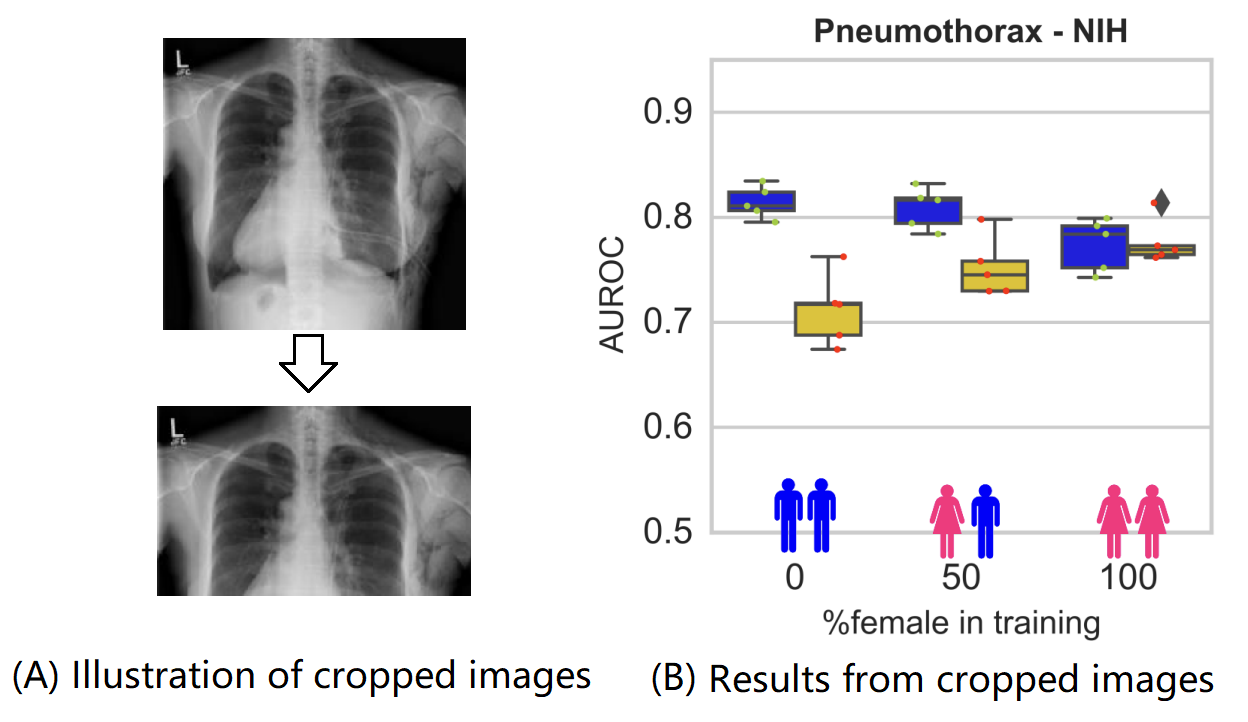}
\caption{The illustration of cropped chest x-rays and the results. 
}
\label{fig:cropped}
\vspace{-0.8cm}
\end{wrapfigure}
To assess specifically whether differences in male and female breast physiology account for the observer performance differences, we perform an additional experiment.
We simply crop the lower two fifths of each recording to ensure that the images contain only the parts above the breast for both genders.
An illustration and the results of this experiment are provided in \cref{fig:cropped}.
Compared to \cref{fig:all_per}, while overall performance drops slightly for both genders, this intervention does not close the performance gap between both genders.

\subsection{Dataset Bias v.s. Model Bias}
\begin{wrapfigure}{r}{0.5\linewidth}
\vspace{-1.8cm}
\centering
\includegraphics[width=\linewidth]{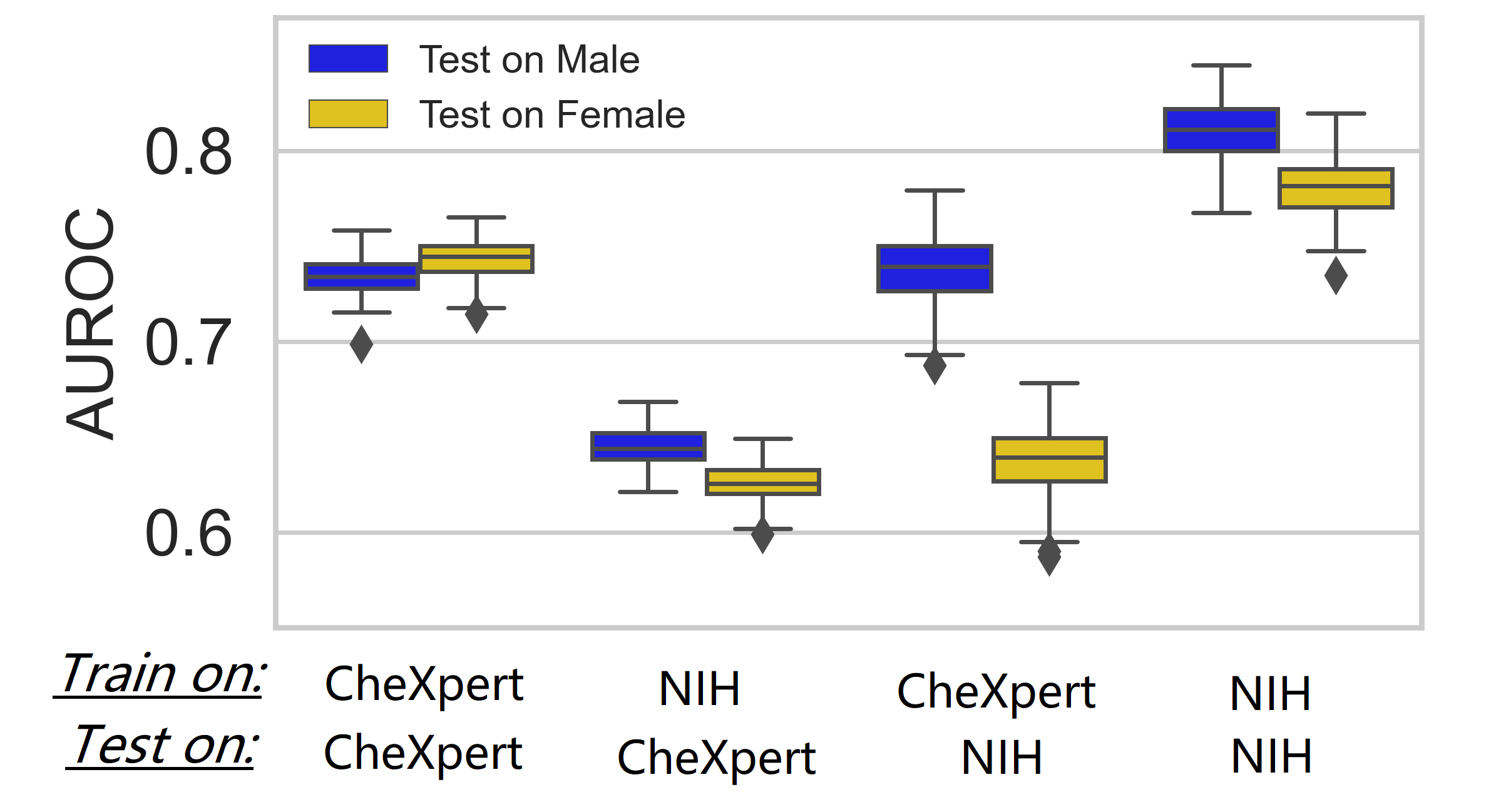}
\vspace{-0.5cm}
\caption{Inter-dataset inference (50\% female in training, \textit{Pneumothorax}) with boost-strapping. 
}
\label{fig:interdataset_res}
\vspace{-1cm}
\end{wrapfigure}
To further investigate the effect of dataset-specific factors, we also evaluated classifiers trained for one disease on CheXpert using NIH test sets, and vice versa (\cref{fig:interdataset_res}).
We observe that large male--female performance gaps only arise when models are tested on NIH, regardless of which dataset the model is trained on.

\section{Discussion \& Conclusions}

We interpret the results of this work and address the key findings as follows.

\paragraph{Supporting evidence of frequent label errors in the ``no finding'' label.}
In agreement with previous reports on other datasets~\cite{zhang2022improving}, our analyses provide supporting evidence of the existing hypothesis on frequent mislabeling of records as ``no finding'' using the proposed simple and easily applicable sampling method.

\paragraph{Male--female performance gaps are influenced, but not fully explained, by training set representation.}
In line with the previous results of Larrazabal et al.~\cite{larrazabal2020gender}, we find that training set representation does (in some diseases strongly) influence male and female model performance.
It does not, however, fully explain the performance gaps: in some cases, even when trained on a fully female dataset, models still perform worse on women (and vice versa).

\paragraph{Male--female performance gaps are not primarily caused by breast shadows.}
It had been previously hypothesized~\cite{ganz2021assessing} that breast shadows might play an essential role in gender bias in ML-based chest X-ray diagnosis, which is not supported by our findings: 
breast cropping does not mitigate the performance gaps. Additionally, the varied bias trends between datasets also contradict this hypothesis.

\paragraph{Biological differences may not be the main driver of male--female performance gaps.}
The performance gaps should be expected to be consistent across datasets (or at least within the same disease), if biological sex differences were the main driver of performance gaps, which has not been observed in this work.

\paragraph{Dataset-specific factors strongly influence male--female performance gaps.}

As the previous hypotheses on the origins of male--female performance gaps have been tentatively rejected considering the results, further research should continue investigating other potential sources of bias. 
Those could be the distribution of various confounders, the prevalence of further label errors, or differing recording quality~\cite{Petersen2023a,Glocker2023,zhang2022improving}.
We conclude that there must be further, at present unknown dataset-specific factors driving the observed performance gaps.

\subsubsection*{Acknowledgements.}
Work on this project was partially funded by the Independent Research Fund Denmark (DFF, grant number 9131-00097B), Denmark’s Pioneer Centre for AI (DNRF grant number P1), a Google Award for inclusion research, the Novo Nordisk Foundation through the Center for Basic Machine Learning Research in Life Science (MLLS, grant number NNF20OC0062606).

%
%
\bibliographystyle{splncs04}
\bibliography{mybibliography}

\end{document}